\documentstyle[12pt]{article}
\input{epsfig.sty}
\newcommand{\beq}{\begin{eqnarray}}
\newcommand{\eeq}{\end{eqnarray}}
\begin{document}
\title{$\Psi$ and $\Upsilon$ Production In pp Collisions at 8.0 TeV}
\author{Leonard S. Kisslinger$^{1)}$\\
Department of Physics, Carnegie Mellon University, Pittsburgh PA 15213 USA.\\
Debasish Das$^{2,3)}$\\
Saha Institute of Nuclear Physics,1/AF, Bidhan Nagar, Kolkata 700064, INDIA.}

\date{}
\maketitle

\begin{abstract}
  This is an extension of recent studies for $\Upsilon(nS), n=1,2,3,$ and 
$J/\Psi(1S),\Psi(2S)$ production at the LHC in pp collisions, and with the 
ALICE detector at 7.0 TeV, with new predictions at 8.0 TeV
\end{abstract}

\noindent
1) kissling$@$andrew.cmu.edu \hspace{1cm} 2)dev.deba$@$gmail.com; 
3) debasish.das@saha.ac.in
\vspace{2mm}

\vspace{2mm}

\noindent
PACS Indices:12.38.Aw,13.60.Le,14.40.Lb,14.40Nd

\section{Differential rapidity cross sections for heavy quark state 
production at ALICE}

  This brief report is a continuation of our work on $\Upsilon(nS)$
production which was published recently\cite{kd13,kd2-13}. It is in anticipation
of the publication of new ALICE experimental results\cite{hxlalice13} on 
$J/\Psi(1S),\Psi(2S)$, and $\Upsilon(nS)$ production in the rapidity range 
$2.5 \leq y \leq 4.0$.

The differential rapidity cross section 
for $\lambda=0$ (dominant for $\Upsilon(nS),\Psi(nS)$ production) is given 
by~\cite{kmm11}
\beq
\label{dsig}
      \frac{d \sigma_{pp\rightarrow \Phi(\lambda=0)}}{dy} &=& 
     A_\Phi \frac{1}{x(y)} f_g(x(y),2m)f_g(a/x(y),2m) \frac{dx}{dy} \; ,
\eeq 
with $a= 4m^2/s$, $s=E^2$, $E=8.0$ TeV, $m=$ 5.0 GeV for Upsilon and 1.5 GeV 
for Charmonium states, $f_g$ the gluonic distribution function, and $x(y),
 \frac{dx}{dy}$ given in Refs\cite{kmm11},\cite{kd13}. For 
Upsilon, Charmonium $a=1.56 \times 10^{-6},1.38 \times 10^{-7}$. $\Phi$ in 
Eq(\ref{dsig}) is either $\Psi$ or $\Upsilon$, with  
$A_\Upsilon=1.33 \times 10^{-8}$ and $A_\Psi=4.95 \times 10^{-7}$.   

The gluonic distribution $f_g(x(y),2m)$ for the range of $x$ needed for $E=8.0$ 
TeV is\cite{kmm11}
\beq
\label{fgupsilon}
         f_g(x(y))&=& 275.14 - 6167.6*x + 36871.3*x^2 \; .
\eeq

  Using the method of QCD sum rules it was shown\cite{lsk09} that
the $\Psi'(2S)$ and $\Upsilon(3S)$ are 50\%-50\% mixtures (with approximately a 
10\% uncertainty) of standard quarkonium and hybrid quarkonium states:
\beq
        |\Psi(2S)>&=& -0.7 |c\bar{c}(2S)>+\sqrt{1-0.5}|c\bar{c}g(2S)>
\nonumber \\
        |\Upsilon(3S)>&=& -0.7 |b\bar{b}(3S)>+\sqrt{1-0.5}|b\bar{b}g(3S)>
 \; ,
\eeq
while the $J/\Psi,\Upsilon(1S),\Upsilon(2S)$ states are essentially standard
$q \bar{q}$ states.
\newpage

  The calculation of the production of $\Upsilon(3S)$ and $\Psi(2S)$ states
is done with the usual quark-antiquark model, and the mixed heavy 
hybrid theory\cite{lsk09}. We find the differential rapidity cross sections 
for $\Upsilon(1S)$, $\Upsilon(2S)$, and $\Upsilon(3S)$ production shown in 
Figure 1.
\vspace{13cm}

\begin{figure}[ht]
\begin{center}
\epsfig{file=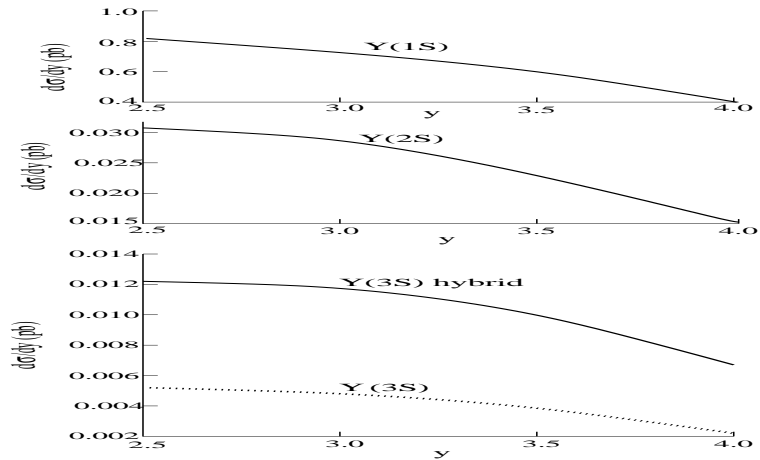,height=6cm,width=10cm}
\end{center}
\caption{d$\sigma$/dy for pp collisions at $\sqrt{s}$ = 8.0 TeV for
producing $\Upsilon(1S)$, $\Upsilon(2S)$; and $\Upsilon(3S)$ for the standard 
model (dashed curve) and the mixed hybrid theory.}
\label{Figure 1}
\end{figure} 

\newpage

The differential rapidity cross sections for $J/\Psi(1S)$ and $\Psi(2S)$  
production for the standard model and the mixed hybrid theory are shown in 
Figure 2.
\vspace{8cm} 

\begin{figure}[ht]
\begin{center}
\epsfig{file=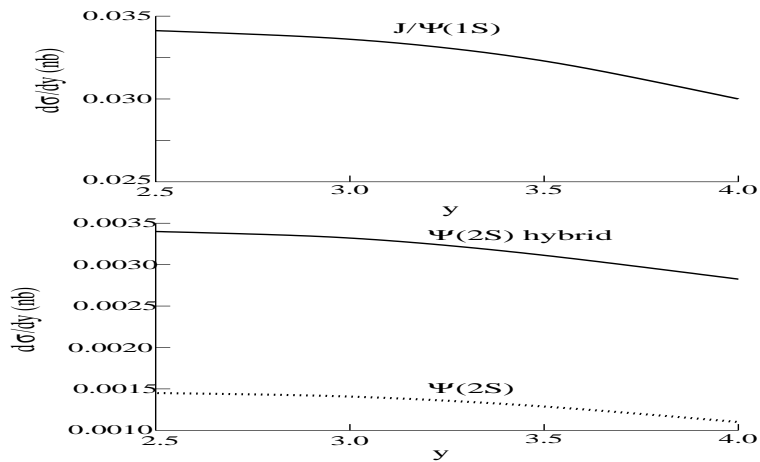,height=6cm,width=10cm}
\end{center}
\caption{d$\sigma$/dy for pp collisions at $\sqrt{s}$ = 8.0 TeV 
producing $J/\Psi(1S)$; and $\Psi(2S)$ for the standard model (dashed curve) 
and the mixed hybrid theory.}
\label{Figure 2}
\end{figure}

\section{Ratios of $\Upsilon(2S),\Upsilon(3S)$  to $\Upsilon(1S)$ 
and $\Psi(2S)$ to $J/\Psi(1S)$ cross sections}

  An essential test of our mixed heavy hybrid theory for heavy quark
production are the ratios of cross section for the production of heavy
quark states. 
\newpage

As discussed in earlier publications\cite{kd13,kmm11} the estimated
$\Upsilon(2S),\Upsilon(3S)$ to $\Upsilon(1S)$ ratios are

\beq
\label{lsk11}
     \Upsilon(2S)/\Upsilon(1S)|_{standard} &\simeq& 
\Upsilon(2S)/\Upsilon(1S)|_{hybrid} \simeq 0.27 \nonumber \\
     \Upsilon(3S)/\Upsilon(1S)|_{standard} &\simeq& .04 \nonumber \\
     \Upsilon(3S)/\Upsilon(1S)|_{hybrid} &\simeq& 0.14-0.22 \; .
\eeq
These ratios have been determined in recent LHCb experiments at 
8 TeV\cite{LHCb13} and 2.76 TeV \cite{LHCb14}. These LHCb measurements show 
that for the $\Upsilon(2S)/\Upsilon(1S)$ ratio the standard model, which 
is the same in the mixed hybrid theory, is correct, while the experimental 
$\Upsilon(3S)/\Upsilon(1S)$ ratio is consistent with the mixed hybrid theory
whereas the standard model prediction is much too small.

From the standard model and  hybrid theory one finds for p-p production of 
$\Psi(2S)$ and $J/\Psi(1S)$
\beq
\label{ppratio}
    \sigma(\Psi(2S))/\sigma(J/\Psi(1S))|_{standard} &\simeq& 0.27 \nonumber \\
    \sigma(\Psi(2S))/\sigma(J/\Psi(1S))|_{hybrid} &\simeq& 0.67\pm 0.07 \; .
\eeq
See Ref. 8 for the recent PHENIX experimental results at forward and central
rapidity for this ratio. Our current predictions will be compared to PHENIX
forward rapidity measurements, which are to be released at the Quark Matter 
2014 Conference in the near future. 

\section{Conclusions}

  We expect that our results for the rapidity dependence of
d$\sigma$/dy shown in the figures, and the ratios of cross sections can be 
useful for experimentalists studying heavy quark production in p-p collisions 
at the LHC.  It is also a test of the validity of the mixed heavy quark hybrid 
theory, for which the ratios of $\Upsilon(3S)/\Upsilon(1S)$ and 
$\sigma(\Psi(2S))/\sigma(J/\Psi(1S))$ have been shown to be in agreement
within errors with the mixed hybrid theory, but not the standard quark-antiquark
model. This is very important since we are using the mixed hybrid heavy quark
theory to test the creation of the Quark Gluon Plasma via Relativistic Heavy 
Ion Collision experiments.

\Large{{\bf Acknowledgements}}

\vspace{5mm}
\normalsize 
Author D.D. acknowledges the facilities of Saha Institute of Nuclear Physics, 
Kolkata, India. Author L.S.K. acknowledges support from the P25 group at Los 
Alamos National laboratory and by a grant from the Pittsburgh Foundation. The
authors acknowledge helpful discussions with members of the PHENIX Collab.
concerning their results which are not yet published.

\end{document}